\input harvmac
\input epsf
\input amssym

\tolerance=10000

%%%%%   version:  12/10/2002

%%%%%%%%%%%%%%%%%%%%%%%%%%%%%%%%%%%%%%%%%
% Basic definitions
%%%%%%%%%%%%%%%%%%%%%%%%%%%%%%%%%%%%%%%%%

%
\def\omit#1{}

\def\coeff#1#2{\relax{\textstyle {#1 \over #2}}\displaystyle}

\def\Tr{{\rm Tr}}
\def\Neq#1{$\cN \!=\! #1$}

 \def\cI{{\cal I}}

\def\cN{{\cal N}} \def\cO{{\cal O}}
\def\cP{{\cal P}} 
\def\cR{{\cal R}} \def\cS{{\cal S}}
 
 \def\cW{{\cal W}}

\def\bfone{\relax{\rm 1\kern-.35em 1}}

%
% Using Blackboard-Bold fontset instead of character kerning
%\def\IC{\relax\,\hbox{$\inbar\kern-.3em{\rm C}$}}
%\def\ID{\relax{\rm I\kern-.18em D}}
%\def\IF{\relax{\rm I\kern-.18em F}}
%\def\IH{\relax{\rm I\kern-.18em H}}
%\def\II{\relax{\rm I\kern-.17em I}}
%\def\IN{\relax{\rm I\kern-.18em N}}
%\def\IP{\relax{\rm I\kern-.18em P}}
%\def\IQ{\relax\,\hbox{$\inbar\kern-.3em{\rm Q}$}}
%\def\IR{\relax{\rm I\kern-.18em R}}
\def\us{\bf}

\def\IR{\Bbb{R}}

\def\eprt#1{{\tt #1}}
\def\nup#1({Nucl.\ Phys.\ $\us {B#1}$\ (}
\def\plt#1({Phys.\ Lett.\ $\us  {#1B}$\ (}
\def\plt#1({Phys.\ Lett.\ $\us  {#1B}$\ (}
\def\cmp#1({Comm.\ Math.\ Phys.\ $\us  {#1}$\ (}
\def\prp#1({Phys.\ Rep.\ $\us  {#1}$\ (}
\def\prl#1({Phys.\ Rev.\ Lett.\ $\us  {#1}$\ (}
\def\prv#1({Phys.\ Rev.\ $\us  {#1}$\ (}
\def\mpl#1({Mod.\ Phys.\ Let.\ $\us  {A#1}$\ (}
\def\ijmp#1({Int.\ J.\ Mod.\ Phys.\ $\us{A#1}$\ (}
\def\atmp#1({Adv.\ Theor.\ Math.\ Phys.\ $\bf {#1}$\ (}
\def\cqg#1({Class.\ Quant.\ Grav.\ $\bf {#1}$\ (}
\def\jag#1({Jour.\ Alg.\ Geom.\ $\us {#1}$\ (}
\def\jhep#1({JHEP $\bf {#1}$\ (}

%

%%%%%%%%%%%%%%%%%%%%%%%%%%%%%%%%%%%%%%%%%

%%%%%

%%%%%%%%%%%%%%%%%%%%%%%%%%%%%%%%%%%%%%%%%
% References
%%%%%%%%%%%%%%%%%%%%%%%%%%%%%%%%%%%%%%%%%
%
%

%\PopeBD
\lref\PopeBD{
C.~N.~Pope and N.~P.~Warner,
``An SU(4) Invariant Compactification Of D = 11 Supergravity On A Stretched Seven Sphere,''
Phys.\ Lett.\ B {\bf 150}, 352 (1985).
%%CITATION = PHLTA,B150,352;%%
}
%
%\PopeJJ
\lref\PopeJJ{
C.~N.~Pope and N.~P.~Warner,
``Two New Classes Of Compactifications Of D = 11 Supergravity,''
Class.\ Quant.\ Grav.\  {\bf 2}, L1 (1985).
%%CITATION = CQGRD,2,L1;%%
}
\lref\CNPNPW{C.~N.~Pope and N.P.~Warner, {\it to appear}.}
%
%\CorradoNV
\lref\CorradoNV{
R.~Corrado, K.~Pilch and N.~P.~Warner,
``An N = 2 supersymmetric membrane flow,''
Nucl.\ Phys.\ B {\bf 629}, 74 (2002)
[arXiv:hep-th/0107220].
%%CITATION = HEP-TH 0107220;%%
}
%
%\CorradoWX
\lref\CorradoWX{
R.~Corrado, M.~Gunaydin, N.~P.~Warner and M.~Zagermann,
``Orbifolds and flows from gauged supergravity,''
Phys.\ Rev.\ D {\bf 65}, 125024 (2002)
[arXiv:hep-th/0203057].
%%CITATION = HEP-TH 0203057;%%
}
%
%\PilchUE
\lref\PilchUE{
K.~Pilch and N.~P.~Warner,
``N = 2 supersymmetric RG flows and the IIB dilaton,''
Nucl.\ Phys.\ B {\bf 594}, 209 (2001)
[arXiv:hep-th/0004063].
%%CITATION = HEP-TH 0004063;%%
}
%
%\FreedmanGK
\lref\FreedmanGK{
D.~Z.~Freedman, S.~S.~Gubser, K.~Pilch and N.~P.~Warner,
``Continuous distributions of D3-branes and gauged supergravity,''
JHEP {\bf 0007}, 038 (2000)
[arXiv:hep-th/9906194].
%%CITATION = HEP-TH 9906194;%%
}
%
%\FreedmanGP
\lref\FreedmanGP{
D.~Z.~Freedman, S.~S.~Gubser, K.~Pilch and N.~P.~Warner,
``Renormalization group flows from holography supersymmetry and a  c-theorem,''
Adv.\ Theor.\ Math.\ Phys.\  {\bf 3}, 363 (1999)
[arXiv:hep-th/9904017].
%%CITATION = HEP-TH 9904017;%%
}
%
%\BuchelCN
\lref\BuchelCN{
A.~Buchel, A.~W.~Peet and J.~Polchinski,
``Gauge dual and noncommutative extension of an N = 2 supergravity  solution,''
Phys.\ Rev.\ D {\bf 63}, 044009 (2001)
[arXiv:hep-th/0008076].
%%CITATION = HEP-TH 0008076;%%
}
%
%\EvansCT
\lref\EvansCT{
N.~Evans, C.~V.~Johnson and M.~Petrini,
``The enhancon and N = 2 gauge theory/gravity RG flows,''
JHEP {\bf 0010}, 022 (2000)
[arXiv:hep-th/0008081].
%%CITATION = HEP-TH 0008081;%%
}
%
%\JohnsonQT
\lref\JohnsonQT{
C.~V.~Johnson, A.~W.~Peet and J.~Polchinski,
``Gauge theory and the excision of repulson singularities,''
Phys.\ Rev.\ D {\bf 61}, 086001 (2000)
[arXiv:hep-th/9911161].
%%CITATION = HEP-TH 9911161;%%
}
%
%\GibbonsZT
\lref\GibbonsZT{
G.~W.~Gibbons and S.~W.~Hawking,
``Gravitational Multi - Instantons,''
Phys.\ Lett.\ B {\bf 78}, 430 (1978).
%%CITATION = PHLTA,B78,430;%%
}
%
%\MaldacenaYY
\lref\MaldacenaYY{
J.~M.~Maldacena and C.~Nunez,
``Towards the large N limit of pure N = 1 super Yang Mills,''
Phys.\ Rev.\ Lett.\  {\bf 86}, 588 (2001)
[arXiv:hep-th/0008001].
%%CITATION = HEP-TH 0008001;%%
}
%
%\GauntlettPS
\lref\GauntlettPS{
J.~P.~Gauntlett, N.~Kim, D.~Martelli and D.~Waldram,
``Wrapped fivebranes and N = 2 super Yang-Mills theory,''
Phys.\ Rev.\ D {\bf 64}, 106008 (2001)
[arXiv:hep-th/0106117].
%%CITATION = HEP-TH 0106117;%%
}
%
%
%\GauntlettSC
\lref\GauntlettSC{
J.~P.~Gauntlett, D.~Martelli, S.~Pakis and D.~Waldram,
``G-structures and wrapped NS5-branes,''
arXiv:hep-th/0205050.
%%CITATION = HEP-TH 0205050;%%
}
%
%\GauntlettRV
\lref\GauntlettRV{
J.~P.~Gauntlett, N.~Kim, S.~Pakis and D.~Waldram,
``M-theory solutions with AdS factors,''
Class.\ Quant.\ Grav.\  {\bf 19}, 3927 (2002)
[arXiv:hep-th/0202184].
%%CITATION = HEP-TH 0202184;%%
}
%
%\GauntlettFZ
\lref\GauntlettFZ{
J.~P.~Gauntlett and S.~Pakis,
``The geometry of D = 11 Killing spinors,''
arXiv:hep-th/0212008.
%%CITATION = HEP-TH 0212008;%%
}
%
%\GunaydinQU
\lref\GunaydinQU{
M.~Gunaydin, L.~J.~Romans and N.~P.~Warner,
``Gauged N=8 Supergravity In Five-Dimensions,''
Phys.\ Lett.\ B {\bf 154}, 268 (1985).
%%CITATION = PHLTA,B154,268;%%
}
%
%\PerniciJU
\lref\PerniciJU{
M.~Pernici, K.~Pilch and P.~van Nieuwenhuizen,
``Gauged N=8 D = 5 Supergravity,''
Nucl.\ Phys.\ B {\bf 259}, 460 (1985).
%%CITATION = NUPHA,B259,460;%%
}
%
%\GunaydinCU
\lref\GunaydinCU{
M.~Gunaydin, L.~J.~Romans and N.~P.~Warner,
``Compact And Noncompact Gauged Supergravity Theories In Five-Dimensions,''
Nucl.\ Phys.\ B {\bf 272}, 598 (1986).
%%CITATION = NUPHA,B272,598;%%
}
%
%
%\SeibergRS
\lref\SeibergRS{
N.~Seiberg and E.~Witten,
``Electric - magnetic duality, monopole condensation, and confinement in N=2 
supersymmetric Yang-Mills theory,''
Nucl.\ Phys.\ B {\bf 426}, 19 (1994)
[Erratum-ibid.\ B {\bf 430}, 485 (1994)]
[arXiv:hep-th/9407087].
%%CITATION = HEP-TH 9407087;%%
}
%
%\SeibergAJ
\lref\SeibergAJ{
N.~Seiberg and E.~Witten,
``Monopoles, duality and chiral symmetry breaking in N=2 supersymmetric QCD,''
Nucl.\ Phys.\ B {\bf 431}, 484 (1994)
[arXiv:hep-th/9408099].
%%CITATION = HEP-TH 9408099;%%
}
%
%\AlvarezGaumeHM
\lref\AlvarezGaumeHM{
L.~Alvarez-Gaume and D.~Z.~Freedman,
``Geometrical Structure And Ultraviolet Finiteness In The Supersymmetric Sigma Model,''
Commun.\ Math.\ Phys.\  {\bf 80}, 443 (1981).
%%CITATION = CMPHA,80,443;%%
}
%
%\SeibergAX
\lref\SeibergAX{
N.~Seiberg,
``Notes on theories with 16 supercharges,''
Nucl.\ Phys.\ Proc.\ Suppl.\  {\bf 67}, 158 (1998)
[arXiv:hep-th/9705117].
%%CITATION = HEP-TH 9705117;%%
}
%
%\CremmerUP
\lref\CremmerUP{
E.~Cremmer and B.~Julia,
``The SO(8) Supergravity,''
Nucl.\ Phys.\ B {\bf 159}, 141 (1979).
%%CITATION = NUPHA,B159,141;%%
}
%
%\deWitIG
\lref\deWitIG{
B.~de Wit and H.~Nicolai,
``N=8 Supergravity,''
Nucl.\ Phys.\ B {\bf 208}, 323 (1982).
%%CITATION = NUPHA,B208,323;%%
}
%
%\deWitNZ
\lref\deWitNZ{
B.~de Wit, H.~Nicolai and N.~P.~Warner,
``The Embedding Of Gauged N=8 Supergravity Into D = 11 Supergravity,''
Nucl.\ Phys.\ B {\bf 255}, 29 (1985).
%%CITATION = NUPHA,B255,29;%%
}
%
%\deWitMZ
\lref\deWitMZ{
B.~de Wit and H.~Nicolai,
``D = 11 Supergravity With Local SU(8) Invariance,''
Nucl.\ Phys.\ B {\bf 274}, 363 (1986).
%%CITATION = NUPHA,B274,363;%%
}
%
%\deWitIY
\lref\deWitIY{
B.~de Wit and H.~Nicolai,
``The Consistency Of The $S^7$  Truncation In D = 11 Supergravity,''
Nucl.\ Phys.\ B {\bf 281}, 211 (1987).
%%CITATION = NUPHA,B281,211;%%
}
%
%\KhavaevGB
\lref\KhavaevGB{
A.~Khavaev and N.~P.~Warner,
``A class of N = 1 supersymmetric RG flows from five-dimensional N = 8  supergravity,''
Phys.\ Lett.\ B {\bf 495}, 215 (2000)
[arXiv:hep-th/0009159].
%%CITATION = HEP-TH 0009159;%%
}
%
%\PilchFU
\lref\PilchFU{
K.~Pilch and N.~P.~Warner,
``N = 1 supersymmetric renormalization group flows from IIB supergravity,''
Adv.\ Theor.\ Math.\ Phys.\  {\bf 4}, 627 (2002)
[arXiv:hep-th/0006066].
%%CITATION = HEP-TH 0006066;%%
}
%
%\KhavaevYG
\lref\KhavaevYG{
A.~Khavaev and N.~P.~Warner,
``A class of N = 1 supersymmetric RG flows from five-dimensional N = 8  supergravity,''
Phys.\ Lett.\ B {\bf 495}, 215 (2000)
[arXiv:hep-th/0009159].
%%CITATION = HEP-TH 0009159;%%
}
%
%\GreeneYA
\lref\GreeneYA{
B.~R.~Greene, A.~D.~Shapere, C.~Vafa and S.~T.~Yau,
``Stringy Cosmic Strings And Noncompact Calabi-Yau Manifolds,''
Nucl.\ Phys.\ B {\bf 337}, 1 (1990).
%%CITATION = NUPHA,B337,1;%%
}
%
%
%\GibbonsVG
\lref\GibbonsVG{
G.~W.~Gibbons, M.~B.~Green and M.~J.~Perry,
``Instantons and Seven-Branes in Type IIB Superstring Theory,''
Phys.\ Lett.\ B {\bf 370}, 37 (1996)
[arXiv:hep-th/9511080].
%%CITATION = HEP-TH 9511080;%%
}
%
%\GibbonsCC
\lref\GibbonsCC{
G.~W.~Gibbons,
``Supergravity Vacua And Solitons,''
% \href{http://www.slac.stanford.edu/spires/find/hep/www?irn=4320514}{SPIRES entry}
{\it Prepared for A Newton Institute Euroconference on Duality and Supersymmetric Theories, Cambridge, England, 7-18 Apr 1997.}
}
%
%\DouglasSW
\lref\DouglasSW{
M.~R.~Douglas and G.~W.~Moore,
``D-branes, Quivers, and {ALE} Instantons,''
\eprt{hep-th/9603167}.
%%CITATION = HEP-TH 9603167;%%
}
%
%\JohnsonPY
\lref\JohnsonPY{
C.~V.~Johnson and R.~C.~Myers,
``Aspects of type {IIB} theory on {ALE} spaces,''
Phys.\ Rev.\ D {\bf 55}, 6382 (1997)
\eprt{hep-th/9610140}.
%%CITATION = HEP-TH 9610140;%%
}
%
%\KachruYS
\lref\KachruYS{
S.~Kachru and E.~Silverstein,
``4d conformal theories and strings on orbifolds,''
Phys.\ Rev.\ Lett.\  {\bf 80}, 4855 (1998)
\eprt{hep-th/9802183}.
%%CITATION = HEP-TH 9802183;%%
}
%
%\KhavaevFB
\lref\KhavaevFB{
A.~Khavaev, K.~Pilch and N.~P.~Warner,
``New vacua of gauged N = 8 supergravity in five dimensions,''
Phys.\ Lett.\ B {\bf 487}, 14 (2000)
[arXiv:hep-th/9812035].
%%CITATION = HEP-TH 9812035;%%
}
%
%%%%%%%%%%%%%%%%%%%%%%%%%%%%%%%%%%%%%%%%%
% Title
%%%%%%%%%%%%%%%%%%%%%%%%%%%%%%%%%%%%%%%%%
\Title{ \vbox{ \hbox{USC-02/05} 
\hbox{\tt hep-th/0212190} }} {\vbox{\vskip -1.0cm
\centerline{\hbox
{Flowing with Eight Supersymmetries}}
\vskip .5cm
\centerline{\hbox { in $M$-Theory and $F$-theory }}
\vskip 8 pt
\centerline{
\hbox{}}}}
\vskip -0.9cm
\centerline{Chethan N.\ Gowdigere and  Nicholas P.\ Warner} 
%%%%%%%%%%%%%%%%%%%%%%%%%%%%%%%%%%%%%%%%%
\bigskip
\centerline{ {\it Department of Physics and Astronomy}} 
\centerline{{\it and}}
\centerline{{\it CIT-USC Center for
Theoretical Physics}}
\centerline{{\it University of Southern California}} 
\centerline{{\it Los Angeles, CA
90089-0484, USA}} 

\vskip 1.0cm
%%%%%%%%%%%%%%%%%%%%%%%%%%%%%%%%%%%%%%%%%
% Abstract
%%%%%%%%%%%%%%%%%%%%%%%%%%%%%%%%%%%%%%%%%
\centerline{{\bf Abstract}}
\medskip
We consider holographic RG flow solutions with eight supersymmetries and 
study the geometry transverse to the brane.  For both $M2$-branes and
for $D3$-branes in $F$-theory this leads to  an eight-manifold with only
a four-form flux.  In both settings there is a natural
four-dimensional hyper-K\"ahler slice that appears on the Coulomb branch.
In the $IIB$ theory this hyper-K\"ahler manifold encodes the Seiberg-Witten
coupling over the Coulomb branch of a $U(1)$ probe theory.  We focus primarily
upon a new flow solution in $M$-theory.   This solution is first obtained using gauged 
supergravity and then  lifted to eleven dimensions.  In this new solution, the brane
probes have an Eguchi-Hanson moduli space with the $M2$-branes
spread over the non-trivial $2$-sphere.  It is also shown that the new solution
is valid for a class of orbifold  theories.  We discuss how the 
hyper-K\"ahler structure
on the slice extends to some form of $G$-structure in the  eight-manifold, and 
describe how this can be computed.

\vskip .3in
\Date{\sl {December, 2002}}
%\draft

%%%%%%%%%%%%%%%%%%%%%%%%%%%%%%%%%%%%%%%%%
% Body
%%%%%%%%%%%%%%%%%%%%%%%%%%%%%%%%%%%%%%%%%
\parskip=4pt plus 15pt minus 1pt
\baselineskip=15pt plus 2pt minus 1pt

%%%%%%%%%%%%%%%%%%%%%%%%%%%%%%%%%%%%%%%%%
\newsec{Introduction}
%%%%%%%%%%%%%%%%%%%%%%%%%%%%%%%%%%%%%%%%%

There  remains a body of important open questions in
the study of holographic flows in AdS/CFT, and
more generally in the study of   string vacua:  
The problem is to  find a geometric characterization 
of {\it supersymmetric } backgrounds that involve
non-trivial fluxes.  There are, of course, well
established theorems for varying levels of
supersymmetry in purely metrical backgrounds, and these
theorems involve hyper-K\"ahler,   K\"ahler and
``Special'' geometry.  There has also been a lot of work on 
supersymmetry breaking using fluxes in Calabi-Yau compactifications.
These solutions often do not incorporate the back-reaction of the branes, or
the flux, or both upon the geometry.  When they do consider  the back-reaction, 
the fluxes are typically arranged so as to yield a  Ricci-flat
background.  Thus such solutions rarely provide exact, large N solutions
with  the AdS space-times that one appears to need for holography.   

For AdS backgrounds, there is  a very significant body of work on the 
classification of solutions 
that involve wrapping $N$ $5$-branes (see, for example, \refs{ \MaldacenaYY,
\GauntlettPS, \GauntlettSC , \GauntlettRV}).  Such backgrounds seem to be amenable
to classifications using $G$-structures, however they are also relatively
simple in that they typically involve only the $NS$ $3$-form flux and the dilaton,
or they involve    the $S$-dual  configuration.
The problem of classifying supersymmetric AdS compactifications in
the presence of multiple $RR$-fluxes, or both $NS$ and $RR$ fluxes, remains wide
open, particularly for compactifications to four and five dimensions.

Perhaps the most vexatious example of this open
problem is the construction of the holographic duals
of ``Seiberg-Witten'' flows.  In particular, one would like to
find the general holographic dual of the flow in which one
gives a mass to a hypermultiplet in the 
large-$N$, $\cN=4$ Yang-Mills theory, and flows to
the Coulomb branch of $\cN=2$, large N, Yang-Mills theory
in the infra-red.  In terms of the $D3$-branes, the Coulomb
branch at large $N$ is represented by a distribution function
for the branes.  On the Coulomb branch of the $\cN=4$ theory,
all six scalars can develop vevs, and the distribution function 
is an arbitrary function of all six variables transverse
to the branes.   It is this function that sources the familiar
harmonic function that underpins these solutions.
 In an $\cN =2$ theory, there are two real 
scalars in the vector multiplet, and so the Coulomb branch
should be represented by an arbitrary function of the two
coordinates corresponding to these scalars.   Crudely speaking,
the complex Higgs invariants, $u_n$, $n=1,\dots,N$ should be the 
coefficients of some form of series expansion of the density 
distribution function of the branes.  Thus the supergravity
dual of the general flow should involve an arbitrary source
function of two variables.

More generally, one would like to know the geometric principle that determines, or
classifies, half-maximal supersymmetry\foot{Here we are taking ``maximal''
to mean $16$ supersymmetries, which is the maximal number in the presence of
a brane.} in the presence of
non-trivial fluxes.  Without fluxes the relevant criterion is
hyper-K\"ahler geometry, which is extremely restrictive  in low dimensions.   
Because the presence of eight supersymmetries so strongly constrains  
the geometry without fluxes, it seems that this amount of supersymmetry
is a reasonable starting point to investigate supersymmetric geometry 
in the presence of fluxes.   

We are by no means going to solve these problems here, but we hope
that this work will assist in its ultimate solution:  One of the
obstructions to solving the problem is the paucity of examples.
To date, there is only one known multiple-flux, ``Seiberg-Witten''
flow solution   \PilchUE, and the corresponding brane distribution
function was determined in \refs{\BuchelCN,\EvansCT}.  Naively this brane distribution
looks like the branes are smeared out uniformly on a disk,
but in reality this ``uniform-disk'' distribution is flattened
onto a line distribution \refs{\BuchelCN,\EvansCT}.
This solution of $IIB$ supergravity was found via the 
methods of gauged $\cN=8$ supergravity in five dimensions 
\refs{\GunaydinQU,\PerniciJU,\GunaydinCU},  and this distribution arises
because it is, in some sense, the most uniform distribution
of branes.  The problem is to generalize this and find the solution 
to $IIB$  supergravity with an arbitrary two-dimensional
distribution function.    The difficulty is that these solutions 
involve non-trivial backgrounds for {\it all} the tensor
gauge fields, as well as for the dilaton and axion of the $IIB$
theory.  The ``simplest'' solution of \PilchUE\  is still far too 
complicated for easy generalizations.    

A useful first step is  to try and recast the solution of
\PilchUE\  in a simpler form.  One way to accomplish this is
to lift the solution to $F$-theory, and then the dilaton
and axion will be encoded in the metric, while the 
$3$-form field strengths are promoted to $4$-forms.
The geometry transverse to the $D3$-branes is then
eight-dimensional, and the most complicated forms
are ``half-dimensional'' on the internal eight-manifold.
This seems very appropriate if one is expecting to generalize
the notion of hyper-K\"ahler.   Indeed,  as we will discuss in section 2, the $F$-theory
lifts of {\it general} Seiberg-Witten flows cleanly encode the dilaton and axion into a
four-dimensional  hyper-K\"ahler slice of the underlying eight-manifold.

A related, and more direct way of trying to generate eight-manifolds associated
with compactifications that have eight supersymmetries  
is to go directly to $M$-theory and study 
flows there.   Such flows are, of course, related by
$T$-duality to flows in the $IIB$ theory, and the geometries
are necessarily very similar (see, for example, \CorradoNV).  
In this paper we will find the $M$-theory analog of
the IIB flow solution of \PilchUE.  It is geometrically
much simpler:  The transverse manifold is eight-dimensional,
and there is only a $4$-form field strength. There is
also a close relationship with the correspondingly
maximally supersymmetric Coulomb flow.    We also find, once again
that there is a four-dimensional hyper-K\"ahler slice analogous to that
of the $F$-theory lift of the $D3$-brane solution. 

To be more precise, we study in detail a two parameter family of
flows.  One parameter involves a scalar vev that preserves an
$SO(4) \times SO(4)$ symmetry, and the other involves turning
on a fermion mass term that breaks this symmetry down to
$(SU(2) \times   U(1))^2$.  A flow with the
scalar vev alone preserves 16 supersymmetries, and generates
a standard Coulomb branch flow that may be described
by the usual harmonic rule with the branes spread out onto
a solid $4$-ball that is rotated by one of the $SO(4)$ factors.
Turning on the fermion mass reduces the number of
supersymmetries to eight.  

As in \refs{\BuchelCN,\EvansCT} we probe our solution with $M2$-branes, and
we find a four-dimensional slice of our solution upon which
the potential of the brane probe vanishes identically:  It is
a slice that corresponds to the Coulomb branch of the theory
on $M2$-branes.   We also find that the metric on this space of
moduli is exactly the Eguchi-Hanson metric, except that there
is a conical singularity at the origin.  When the fermion mass is non-zero, the branes that 
were  spread on a solid $4$-ball on the pure Coulomb branch, are
localized on the non-trivial $2$-sphere of the Eguchi-Hanson
space.  We thus obtain a beautifully simple hyper-K\"ahler section 
through our solution.   We also discuss
a more general class of orbifold flows that have eight supersymmetries,
and for which the brane-probe moduli space is the Eguchi-Hanson space 
everywhere, with no singularities. 

There are thus several purposes to this paper.  First, we find  and elucidate 
a relatively simple flow geometry with  eight supersymmetries.  There is only
one other explicitly known example in the same class, and so we are in
this sense  doubling the number of examples. We have
also found the residual hyper-K\"ahler geometry in this solution, and
in its type $IIB$ bretheren.   In section 2 we will describe how such geometry
arises naturally in the $F$-theory lifts of Seiberg-Witten geometries, and 
how it appears implicitly in the results of \refs{\BuchelCN,\EvansCT}.  We 
also discuss how it
is related to the hyper-K\"ahler geometry of the moduli space
of $M2$-brane probes.  

In section 3 we start the discussion of our new class of flows.
We approach this problem in the same manner as \refs{\PilchUE,
\FreedmanGK,\FreedmanGP}:
We first obtain  the flows in gauged \Neq8 supergravity
in four dimensions.  In a small aside, we conclude section 3 by noting that
our results show that there is a two parameter family of solutions to $M$-theory
in which the  parameters trade metric deformations for background fluxes.
Indeed, when the parameters are zero, the flow is purely the harmonic
metric perturbation corresponding to a Coulomb branch flow.  Changing the 
parameters smoothly turns on background fluxes.    Members of
the family of flows generically have no supersymmetry, but there is a one parameter 
family that preserves eight supersymmetries, and another that preserves only 
four supersymmetries.

Those who do not care for the details of $\cN=8$ supergravity should skip to
section 4 where we give the corresponding solution in eleven dimensions.
We also show that our new solution is a very
close,  $M$-theory analog of the solution of \PilchUE.  In  particular we
examine the geometric symmetries and their actions on
the supersymmetries, and identify the two directions within the transverse
eight-geometry that represent the two ``added''  directions 
when compared to the transverse six-geometry of the $D3$-brane
flow given in \PilchUE.   We also discuss orbifolds of our solution that 
preserve all eight supersymmetries.

 In section 5 we first look at the pure Coulomb 
branch flow (the one with 16 supersymmetries) and exhibit its canonical 
harmonic form.  We then use $M2$-brane probes and find the loci
upon which they feel ``zero force.''  One of these loci corresponds to
the directions in which the branes spread on the Coulomb branch, but when the
fermion mass is turned on we show that the metric on the space in 
the Eguchi-Hanson metric, and that the branes 
localize on the non-trivial $2$-sphere of the Eguchi-Hanson.

Finally, section 6 contains our conclusions as well as some discussion about
how we expect to be able to generalize the hyper-K\"ahler forms
out into the full eight-geometry of our solution.

%%%%%%%%%%%%%%%%%%%%%%%%%%%%%%%%%%%%%%%%%
\newsec{Yang-Mills couplings  and Hyper-K\"ahler moduli spaces}
%%%%%%%%%%%%%%%%%%%%%%%%%%%%%%%%%%%%%%%%%

It follows from the work of \AlvarezGaumeHM\ that an $\cN =4$ supersymmetric sigma 
model  must be hyper-K\"ahler.  In particular, given a perturbative 
$M2$-brane background that preserves eight supersymmetries
(and is thus 
$\cN =4$ supersymmetric on the world volume), the
transverse coordinates of 
the brane must give rise to a 
hyper-K\"ahler sigma-model on the 
world-volume.  There is a possible complication at large $N$ in that the 
theory on the $M2$-brane is conformal, strongly coupled, and there is no 
known explicit description of the field theory.  
On the other hand one can still use brane probes, and if there is a
zero-force domain for the probe, then the corresponding moduli
space will necessarily be hyper-K\"ahler.

For $D3$-branes in $IIB$ supergravity, eight supersymmetries
mean $\cN=2$ supersymmetry and thus {\it a priori} one merely has
an ordinary K\"ahler moduli space. However, if one rewrites this as an 
$F$-theory compactification, then the inclusion of the elliptic fiber, $T$, of 
$F$-theory will typically  promote the K\"ahler moduli space to a 
hyper-K\"ahler space.  This does not follow directly from the $IIB$ picture since there 
is not enough supersymmetry on the $D3$-brane, and there are no sigma
model dynamics in the elliptic fiber.  However, one can establish
this result using duality:  Wrap one of the $D3$-brane directions around a
circle, $C$, and $T$-dualize.  The resulting $IIA$ background
can then be thought of as $M$-theory compactified on a torus with
the same complex structure as $T$, but whose K\"ahler modulus is 
the radius of $C$.  In the $M$-theory, the moduli space
of the $M2$-branes is the torus $T$ fibered over the original 
moduli space of the $D3$-branes.  Moreover, since the $M$-theory
solution was obtained as the dual of $D3$-branes, the $M2$ branes
will be ``smeared'' over the circles of the torus. 

One can see this very naturally in the $D3$-brane description of
$\cN =2^*$ supersymmetric flows from $\cN=4$ Yang-Mills.
In such flows the space transverse to the branes is topologically
$\IR^6$, with the coordinates split into the $2+4$ corresponding to
the scalars of the $\cN=2$ vector and hypermultiplets respectively.  
Let $z$ be a complex coordinate corresponding to the former, so that displacing
the branes in the $z$-direction corresponds to deforming the theory
out onto the $\cN=2$ Coulomb branch.   A brane probe will therefore
experience no force if it approaches the solution along
the $z$-plane through the origin, and the value of the $IIB$ dilaton/axion, 
$\tau(z)$, is the complex gauge coupling of the $U(1)$ on the probe.  
Because of the usual properties of $\cN=2$ Yang-Mills, $\tau(z)$ 
is a holomorphic function.

Let $ds_2^2$ be the metric felt by the probe on the
$\cN=2$ Coulomb branch.  The scalar kinetic term for the Yang-Mills
bosons   provides the canonical metric on the moduli space of the probe:
\eqn\natmet{ ds_2^2  ~=~  \tau_2 \, |dz|^2 ~=~ 
e^{-\phi} \, |dz|^2 \,,\ \  {\rm where} \ \ 
\tau = \tau_1 + i \tau_2 = {\theta_s \over 2 \pi} + 
{i \over g_s} \,.}
It is, of course guaranteed within the field theory that $\cI m(\tau) >0$,  
and indeed this is one of the essential  ingredients in obtaining the Seiberg-Witten 
effective action \refs{\SeibergRS,\SeibergAJ}.  

Within $F$-theory, modular invariance on the elliptic fiber fixes
the form of the fiber metric, and so the   $4$-metric  of the fibration over 
\natmet\ must be:
\eqn\HKmet{ ds_4^2  ~=~ {1 \over \tau_2} \, | d  \varphi_1 + 
\tau \, d \varphi_2 |^2 ~+~ \tau_2 \, |dz|^2 ~=~
{1 \over \tau_2} \, ( d \varphi_1 + \tau_1 \, d \varphi_2)^2 ~+~ 
\tau_2\, \big( d\varphi_2^2 +  dx^2 +  dy^2 \big)   \,,}
where $z= x+ iy$.  This is precisely the kind of metric that was
considered in the early investigation of stringy cosmic strings 
\refs{\GreeneYA,\GibbonsVG,\GibbonsCC}.
Remember that $\tau = \tau(z)$ is holomorphic,
and thus $\tau_1(x,y)$ and $\tau_2(x,y)$ are harmonic functions that
satisfy the Cauchy-Riemann equations.  Using this one can easily
verify that \HKmet\ is Ricci flat, with a anti-self-dual curvature tensor.
That is, \HKmet\ is a ``half-flat'' $4$-metric with $SU(2)$ holonomy, and 
is thus hyper-K\"ahler.  There are two covariantly constant spinors
of the same helicity, and 
the three closed forms that make up the hyper-K\"ahler structure are:
\eqn\HKforms{\eqalign{ J  ~=~ & \coeff{i}{2\, \tau_2} \, (d \varphi_1 + 
\tau \, d\varphi_2) \wedge  (d \varphi_1 +  \bar \tau \, d\varphi_2)  ~+~ 
\coeff{i\, \tau_2}{2}  \, dz \wedge d \bar z ~=~d \varphi_1 \wedge 
d\varphi_2 ~+~ \tau_2 \, dx \wedge dy \cr 
\Omega  ~=~ & (d \varphi_1 +  \tau \, d\varphi_2) \wedge dz \,, 
\qquad \qquad  \overline{\Omega}  ~=~  (d \varphi_1 + 
\bar \tau \, d\varphi_2) \wedge d\bar z  \,.}}

The metric \HKmet\ has a very similar form to the Gibbons-Hawking
ALE metrics \GibbonsZT:
\eqn\ALEmet{ ds^2  ~=~ V^{-1}\, (d \varphi + A)^2 ~+~ V (\, d\vec x \cdot 
d\vec x \,)  \,, }
where
\eqn\VAforms{V(x) ~\equiv~ 2\,m\, \sum_{j=1}^M \, {1 \over |\vec x
- \vec x_j|} \,, \qquad \vec \nabla \times \vec A ~=~ \vec\nabla V\,. }
Indeed, one may think of \HKmet\ as being exactly of this 
form, but with the point sources in the potential being replaced by
line-sources wrapped around the compactified $x_3$ direction. 
The harmonic potential and vector field are then  $V(x) = \tau_2, 
\vec A =(\tau_1, 0,0)$, and the relationship between $V$ and
$\vec A$ reduces to  the Cauchy-Riemann equations.

Following the discussion of \BuchelCN, recall that 
in terms of the field theory, the probe and the original $N$ 
$D3$-branes  represent the breaking of $SU(N+1) \rightarrow 
U(1) \times U(1)^{N-1}$.  On the Coulomb branch, the complex scalar 
field in the vector multiplet develops the generic vev:
\eqn\Phivev{\Phi~=~ {\rm diag}\big(z, a_1 -{z \over N}, a_2 -{z \over N}, 
\dots, a_N -{z \over N}\big) \,,\qquad \sum_{j=1}^N a_j =0 \,.}
If one computes $\tau(z)$ for the Wilsonian effective action as in
\refs{\SeibergRS,\SeibergAJ}, $\tau(z)$ is holomorphic, with $\cI m(\tau) >0$, 
and so the metric \HKmet\ will be non-singular, except when there are 
massless BPS states, which  appear precisely when the probe brane
encounters any of the other ``fixed''  branes.   
 
In the large $N$ limit, the effective action of the $\cN=2^*$ 
gauge theory reduces to its perturbative, one loop form, and so
\eqn\perttau{\tau(z) ~=~ {i \over g_s} ~+~ {\theta_s \over 2 \,\pi} ~+~
{i \over 2\,\pi} \, \sum_{j=1}^N \,\ln\bigg[ {(z-a_j+ {z \over N})^2 \over
(z-a_j+ {z \over N})^2 -m^2 } \bigg]  \,.}
However, in AdS/CFT the discrete set of branes at points $a_j$ will 
be also replaced by a continuum distribution.  In particular,  for the $D3$ brane
flow of \PilchUE\ it was shown in \refs{\BuchelCN,\EvansCT} that:
\eqn\specperttau{\tau(z) ~=~ {i \over g_s} \, \bigg({z^2 \over z^2 - a_0^2} 
\bigg)^{1 \over 2}  ~+~ {\theta_s \over 2 \,\pi}   \,,}
corresponding to a linear brane density distribution with density:
\eqn\branedensity{\rho(x) ~=~\sqrt{a_0^2- x^2}  \,,}
along the $x = \cR e(z)$ axis between $-a_0$ and $a_0$.

Because \perttau\ and \specperttau\ are merely perturbative forms
of $\tau$, the imaginary part of  $\tau$ can vanish at finite values of $u$, 
and the metric \HKmet\ will be singular at such points.    For the solution
with \specperttau, one has $\cI m (\tau)=0$ along the $ \cR e(z)$ axis between 
$-a_0$ and $a_0$, which is precisely where the branes are located.  Thus one
is really dealing with singular moduli spaces from which the brane sources
will need to be excised.  Indeed,  it has been suggested that such singularities 
can be resolved in  the AdS/CFT correspondence through the enhan\c con mechanism
in which one excises the interior of the region where $\cI m (\tau)$ vanishes, and 
replaces this region by a flat parameter space.

In this paper we will be focusing on $M2$-brane flows, and so we expect
that the brane distributions will not be smeared around circles of compactification,
but will be localized to the four-dimensional space of moduli.  Indeed, if
the  moduli space is non-trivial then one should expect that it will have the form
given by \ALEmet\ and \VAforms, and this is indeed what we find.

%%%%%%%%%%%%%%%%%%%%%%%%%%%%%%%%%%%%%%%%%
\newsec{Flows in four-dimensional supergravity}
%%%%%%%%%%%%%%%%%%%%%%%%%%%%%%%%%%%%%%%%%

We will describe some holographic flows in the maximally
supersymmetric field theory on a stack of $(N+1)$ $M2$-branes.
There are eight dimensions transverse to the branes, and
so the $\cR$-symmetry is $SO(8)$.  The theory on a single brane 
consists of eight bosons, $X^I$, transforming in the $8_v$
of $SO(8)$, and eight fermions, $\lambda^a$, 
transforming in the $8_c$.   This theory has eight supersymmetries,
$Q^{a'}$, transforming in the $8_s$, and acting through
triality.  On a stack of $N+1$ such branes there are
$8(N+1)$ such bosons and fermions corresponding to the supermultiplet
of the transverse locations of the branes.  Perturbatively, these
would correspond to scalars in the Cartan subalgebra of
the underlying $SU(N+1)$ gauge group, but the superconformal
theory only emerges in a strong coupling limit \SeibergAX.   More precisely,
The field theory corresponding to the center of mass motion remains
a free theory, but the other degrees of freedom have an
infra-red, \Neq8 superconformal fixed point at which the 
coupling goes to infinity.
This field theory also arises as a  Kaluza-Klein reduction of 
\Neq4 supersymmetric Yang-Mills theory on
a circle. The extra scalars in three-dimensions come from the Wilson line
parameter around the circle and from dualising the three-dimensional 
photon.

Gauged \Neq8 supergravity in four dimensions contains 
70 scalar fields, and these are holographically dual to operators
that have a perturbative form of  (traceless) bilinears in the 
scalars and fermions:
\eqn\bilinears{\eqalign{\cO^{IJ} ~=~ & \Tr~\big(X^I \, X^J) ~-~ 
\coeff{1}{8}\, \delta^{IJ}\, \Tr~\big(X^K \, X^K\big) \,, 
\quad I,J, \dots =1,\dots,8\cr \cP^{AB} ~=~ & \Tr~\big(
\lambda^A \, \lambda^B\big) ~-~ \coeff{1}{8}\, \delta^{AB}\, 
\Tr~\big(\lambda^C \, \lambda^C \big) \,, \quad A,B,
\dots =1,\dots,8\,,}}
where $\cO^{IJ}$ transforms in the ${\bf 35}_v$  of $SO(8)$, and 
$\cP^{AB}$  transforms in the ${\bf 35}_c$.
Thus,  gauged  \Neq8  supergravity in four dimensions can be used to study
mass perturbations, and a uniform subsector of the  Coulomb branch of
the \Neq8  field theory on the large-$N$ stack of $M2$-branes.  The gauged
\Neq8  supergravity in four dimensions thus plays a very analogous
role to the  gauged \Neq8 supergravity in five dimensions.
There is, however, a significant difference:  the Yang-Mills theory
has a freely choosable (dimensionless) coupling constant and
$\theta$-angle, and these are dual to a pair of scalars
in the five-dimensional gauged supergravity theory.  The scalar-fermion
theory on the $M2$ branes has no free coupling: The perturbative
gauge coupling flows to infinity in the fixed point theory. 
There are thus no supergravity fields dual to a 
coupling: there are only masses and vevs in the dual of 
the four-dimensional  gauged supergravity.

\subsec{A simple family of flows}

We wish to consider flows with a pair of
independent fermion masses and a  pair of
independent boson masses.  We would like them
to be parallel to the  two-mass flows in four dimensions
in which the fermion masses are $m_1 \lambda^3
\lambda^3 + m_2 \lambda^4 \lambda^4$.
In the three-dimensional theory we therefore want to 
consider perturbations involving operators of the form:
\eqn\pertOps{\eqalign{\cO_1 ~\equiv~ a_1\, \big(\cO^{11} + 
\cO^{22} + \cO^{33} +  \cO^{44} \big) ~+~  a_2\, 
\big(\cO^{55} + \cO^{66}  \big)  ~+~ a_3\, \big(\cO^{77} + 
\cO^{88} \big) \,, \cr \cO_2 ~\equiv~ b_1\, \big(\cP^{11} + 
\cP^{22} + \cP^{33} +  \cP^{44} \big) ~+~  b_2\, 
\big(\cP^{55} + \cP^{66}  \big)  ~+~ b_3\, \big(\cP^{77} + 
\cP^{88} \big) \,,  }}
where tracelessness requires $a_1 + a_2 + a_3 =0$ and
$b_1 + b_2 + b_3 =0$.

In the  half-maximal supersymmetric flow
of \PilchUE\ in four dimensions the perturbing operators
were:
\eqn\twoflowops{ \cO_b ~=~ \sum_{j=1}^4\,  {\rm Tr} \big( X^j
X^j \big)  \,-\, 2\sum_{j=5}^6 {\rm Tr}\big( X^{j} X^{j})   \,, \qquad
\cO_f ~=~ {\rm Tr} \big( \lambda^3 \lambda^3 + \lambda^4
\lambda^4 \big)  \,,}
and the analogues of these operators on the $M2$-brane
are given by \pertOps, with $a_2=a_3 =-a_1$, 
$b_1=0, b_2 =-b_3$.

Since we wish to focus upon flows that involve
very particular operators we need to truncate the 
70 scalars to a  consistent subsector.
A simple way to accomplish this is to truncate to
the singlet sector of a carefully chosen symmetry.
Each of the operators in \pertOps\ is invariant
under $SO(4) \times SO(2) \times SO(2)$, but the
$SO(4)$'s are not the same since the indices in
$\cO_1$ and $\cO_2$ are $SO(8)$ vectors and 
spinors respectively.  The common invariance 
group is thus $H \equiv SU(2) \times (U(1))^3$, and we will
ultimately focus on the flows with this invariance
group.

\subsec{Another approach to the Ansatz}

It is useful to consider the other manner in which 
we might have generalized the results of \PilchUE.  Recall that this 
solution is invariant under $SU(2) 
\times  U(1) \subset SO(4) \subset SO(6)$.  Another natural
generalization is to consider the scalar manifold with 
precisely the same kind of invariance.   That is,
consider the scalars  invarant under $H_0 \equiv SU(2) 
\times  U(1)$, where the $U(1)$ lies in one of the $SU(2)$'s
in $(SU(2))^4 \subset SO(8)$.    Within $SO(8)$, the group $H_0$
commutes with $SU(2) \times SU(2) \times U(1)$, where
this $U(1)$ is the same  as the $U(1)$ that appears in $H_0$.
We will drop this $U(1)$ in the commutant henceforth.
Within $SU(8)$, $H_0$ commutes with $SO(6) \times U(1)$
and within $E_{7(7)}$ it commutes with $SO(6,1) \times
SL(2,\IR)$.  Conversely, this latter group commutes with
$SO(5)$, and so $H_0$ is the $SO(3) \times SO(2)$ subgroup
of $SO(5)$ that commutes with $SO(6,1) \times
SL(2,\IR)$ in $E_{7(7)}$.  The invariant scalar manifold
is therefore:
\eqn\scalman{\cS_0 ~\equiv~ {SO(6,1) \over SO(6)} \times
{SL(2,\IR)  \over SO(2)}  \,.}
There are thus, {\it a priori}, eight scalars.
Six of them may be thought of as being an $SO(6)$
vector, $\vec v$.
However,  the residue of the $SO(8)$ symmetry acts on $\cS$
as $SO(3) \times SO(3) \subset SO(6)$, and this
may be used to reduce $\vec v$ to the form
$(v_1,0,0,v_4,0,0)$, which has two parameters and
is manifestly invariant under $SO(2) \times SO(2) 
\subset SO(3) \times SO(3)$.  This extends $H_0$
to $H = SU(2) \times (U(1))^3$, and indeed this special
``gauge'' has reduced the apparently more general scalar 
manifold, $\cS$, to the simpler one:
\eqn\redscalman{\cS ~\equiv~ {SL(2,\IR)  \over SO(2)} \times
{SL(2,\IR)  \over SO(2)}  \,.}
that is invariant under $H$ and
whose scalars are dual to the operators \pertOps.
Either generalization yields the same end result.

For future reference, it is useful to note
that each $SL(2,\IR)$ contains two supergravity
scalars: one dual to a scalar bilinear in $\cO_1$, and the
other dual to a scalar bilinear in $\cO_2$.  In  terms
of group theory: The non-compact  generators in
each of the $SL(2,\IR)$'s  consist of one from the 
$35_v$ of $SO(8)$ and the other from the $35_c$.  The compact
generators come from  the $35_s$ that extends $SO(8)$ to
$SU(8)$, and so the compact generators are not, 
{\it a priori}, symmetries of the theory.
 
\subsec{The flows in gauged supergravity}

 Following \refs{\CremmerUP,\deWitIG}, define the action of the  $E_{7(7)}$ by:
\eqn\Evars{\eqalign{ \delta\, z_{IJ} ~=~ \Sigma_{IJKL}\,
z^{KL} \cr \delta\, z^{IJ} ~=~ \Sigma^{IJKL}\,
z_{KL} \,.}}
where indices are raised and lowered by complex 
conjugation, and where one has:
$$
\Sigma_{IJKL} ~=~  \overline{(\Sigma^{IJKL})}
~=~ \coeff{1}{24}\, \varepsilon_{IJKLPQRS}\, 
\Sigma^{PQRS} \,.
$$
Introduce a complex, skew, self-dual, tensor 
$X_{ab} = - X_{ba}$, $a,b =1,\dots,4$, by setting:
\eqn\Xdefn{X_{12}  ~=~ \overline X_{34}   ~=~ z_1\,, \qquad
X_{13} ~=~ - \overline  X_{24} ~=~ z_2\,, \qquad 
X_{23} ~=~ \overline  X_{14} ~=~ z_3\,.}
The non-compact  generators of $SO(6,1) \times SL(2,\IR)$ 
may then be written as:
\eqn\GPgens{ \Sigma_{IJKL} ~=~ 24 \, \big(\, z_0 \, 
\delta^{1234}_{[IJKL]} ~+~
\bar z_0\,\delta^{5678}_{[IJKL]} \,\big) ~+~  24 \, 
X_{ab} \, \big( \delta^1_{[\, I} \delta^2_J  ~+~ 
\delta^3_{[\, I} \,\delta^4_J  \big) \,
\delta^{a+4}_K \, \delta^{b+4}_{L\,]} 
\,,}
where $z_0$ parametrizes the $SL(2,\IR)$, and
$z_j$, $j=1,2,3$ parametrizes
the non-compact generators of $SO(6,1)$.   The 
gauge choice  that reduces the scalar
manifold to \redscalman\ is to set $z_2 =z_3 =0$.
The real and imaginary parts of $z_j = x_j + i y_j$,
$j=0,\dots,3$ correspond to the $35_v$ and $35_c$
and so parametrize the operators $\cO_1$ and
$\cO_2$ respectively.

To get a sense of what these generators represent, one can take
$y_j =0$ and contract the $\Sigma$'s  with suitably chosen gammma 
matrices, and define:
$$
S_{AB} ~\equiv~ \Sigma_{IJKL} \, \big(\Gamma^{IJKL} \big)_{AB} \,.
$$
One then finds that $S$ is diagonal, and of the form:
\eqn\Sdiag{ S ~=~ {\rm diag} (-\alpha,\, -\alpha,\, -\alpha,\, 
-\alpha, \,  \alpha - 2\,\chi, \,  \alpha - 2\,\chi, \,
\alpha + 2\,\chi, \, \alpha + 2\,\chi) \,,}
which shows how these scalars map onto \pertOps.

It is straightforward to exponentiate
the actions, \Evars, of these group generators
and construct the $56 \times 56$ matrices
that underlie the $\cN=8$ supergravity in
four dimensions.  
One can then assemble the potential and
$SU(8)$ tensors that are central to the
structure of the gauged $\cN=8$ theory.

We will use polar coordinates and take:
\eqn\polcoords{z_0 = \coeff{1}{2}\, \alpha \, e^{ i \,\phi}\,, 
\qquad z_1 = \coeff{1}{2}\, \chi \,e^{ i \,\varphi}\,, 
\qquad z_2 =z_3 =0 \,.}
If the angles, $\phi$ and $\varphi$, vanish then the
supergravity scalars are in the $35_v$ and correspond
to the operators in $\cO_1$, and if $\phi= \varphi ={\pi \over 2}$, 
then the supergravity scalars are in the $35_c$ and correspond
to the operators in $\cO_2$.

One  finds that the scalar kinetic term has the form:
\eqn\scalkin{ (\partial_\mu \, \alpha)^2 ~+~ \coeff{1}{4}\,  
\sinh^2(2\,\alpha)\, (\partial_\mu \, \phi)^2~+~ 
2\,\big[ (\partial_\mu \, \chi)^2 ~+~ \coeff{1}{4}\,  
\sinh^2(2\,\chi)\, (\partial_\mu \, \varphi)^2 \big]\,,}
which determines the canonically normalized scalars to
be $\beta_1 =  2\, \chi $ and $\beta_2 = \sqrt{2}\, \alpha$.

Rather surprisingly (given that the compact generators
of \redscalman\ are not a symmetry of the overall theory)
one finds that the supergravity potential is completely
independent of the angles $\varphi$ and $\phi$, and is given
by the simple formula:
\eqn\sugrpot{\cP ~=~ -{1\over L^2} \, (\cosh(2\,\alpha) ~+~ 
2\, \cosh(2\,\chi)) \,,}
where we have replaced the usual supergravity gauge coupling,
$g$, according to:
\eqn\gLreln{ g~=~ {1 \over \sqrt{2} \, L} \,.}
With these conventions, the maximally symmetric critical point
at $\alpha = \chi =0$ gives rise to an AdS$_4$ vacuum of 
radius $L$.

The $SU(8)$ tensor, $A_1^{ij}$, that appears in the gravitino
variation is real and has  constant eigenvectors with eigenvalues:
\eqn\candWs{\eqalign{\cW_1 ~=~& \cosh (\alpha)\, \cosh^2(\chi) ~+~ 
e^{-i\,\phi} \, \sinh (\alpha)\, \sinh^2(\chi) \,, \cr
\cW_2 ~=~& \cosh (\alpha)\, \cosh^2(\chi) ~+~ e^{-i\,(2\, 
\varphi - \phi)} \, \sinh (\alpha)\, \sinh^2(\chi) \,, \cr
\cW_3 ~=~& \cosh (\alpha)\, \cosh^2(\chi) ~+~ e^{ i\,(2\, 
\varphi + \phi)} \, \sinh (\alpha)\, \sinh^2(\chi) \,,}}
with multiplicities $4,2$ and $2$ respectively.  The constancy
of the eigenvectors makes these eigenvalues good candidates
for superpotentials \KhavaevGB.

One can easily verify that all of these eigenvalues are
related to the potential, $\cP$, via:
\eqn\PWreln{\cP ~=~ {1\over L^2}\, \bigg|{\del \cW \over \del 
\alpha}  \bigg|^2 ~+~ {1\over 2\, L^2}\, \bigg|{\del \cW \over 
\del \chi}  \bigg|^2 ~-~  {3\over L^2}\, |\cW |^2 \,.}
The absence of derivatives with respect to $\phi$
and $\varphi$ here means that the $\cW_j$ will only
be superpotentials if the angles are fixed in a manner
consistent with supersymmetry.   

As is conventional, we will consider a flow metric of the
form:
\eqn\RGFmetric{
ds^2_{1,3} = e^{2 A(r)} \eta_{\mu\nu} dx^\mu dx^\nu + dr^2 \,,}
where $\eta_{\mu\nu}$ is the flat, Poincar\'e invariant metric
of the $M2$-brane.  The supersymmetric flow equations
are then:
\eqn\floweqs{ {d \alpha \over d r} ~=~  - {1 \over L}\,
{\del \cW \over \del \alpha}  \,, \qquad 
{d \chi \over d r}  ~=~ -  {1 \over 2\, L}\, {\del \cW \over
\del \chi}  \,, \qquad
{d A \over d r} ~=~ {1 \over L} \, \cW  \,.} 

There are several natural choices of superpotential.  The simplest
is to take $\phi = \varphi=0$ and then all the $\cW_j$
are equal.  The corresponding flows involve only the operators
in $\cO_1$, preserve \Neq8 supersymmetry, and represent pure
``Coulomb branch'' flows from the $UV$ fixed point theory.

The flow that we will focus on here comes from taking:
\eqn\angchoice{ \phi = 0\,, \qquad \varphi={\pi \over 2}\,.}
From \Sdiag\ we then see that $z_0$ preserves $SO(4) \times
SO(4)$, and then $z_1$ breaks this down to $(SU(2) \times U(1))^2$.
For this family of scalars, the superpotential is given by:
\eqn\superpot{\eqalign{\cW ~=~ \cW_1 ~=~ & \cosh (\alpha)\, 
\cosh^2(\chi) ~+~    \sinh (\alpha)\, 
\sinh^2(\chi) \cr ~=~ & \coeff{1}{2}\, \big(\rho^{-1} ~+~
\rho \, \cosh(2\, \chi) \big)\,,}}
where $\rho \equiv e^\alpha$.
We could equally well have used $\cW= \cW_2 = \cW_3$, which is
related to $\cW_1$ by $\alpha \to -\alpha$.
The important point is that with the choice \angchoice\ 
there are two sets of four distinct eigenvalues of
$A_1^{ij}$, and hence the flows are going to be
$\cN=4$ supersymmetric; that is, the flows will have
half-maximal supersymmetry.  Indeed, the four unbroken supersymmetries
 are singlets under one $SU(2) \times U(1)$ factor, and transform as a 
${\bf 2}_{\pm 1}$ of the other $SU(2) \times U(1)$. 

In terms of the dual theory on the brane, one can see from \Sdiag, \polcoords\
and \angchoice\ that these flows involve the operators
$\cO_1$ and $\cO_2$ with $a_2 = a_3 = - a_1$; $ b_1 =0,
b_2 = - b_3$.  As noted after \twoflowops, this is indeed
the precise analogue of the flow in \PilchUE.

As in \PilchUE, one can solve the flow equations completely,
and one finds:
\eqn\alsoln{\alpha ~=~ \coeff{1}{2} \, \log\big[ 
e^{- 2 \, \chi} ~+~ \gamma\, \sinh(2\, \chi) \big] \,,}
with
\eqn\Asoln{e^A  ~=~ {k \, \rho \over \sinh(2\,\chi)} \,,}
where $k$ and $\gamma$ are constants of integration.
The solution \alsoln\ is considerably simpler than the
analogue in \PilchUE, and indeed there are some interesting
special cases with $\alpha = \pm \chi$ and $\rho^2
=\cosh(2\chi)$.    
The solution \alsoln\  has different asymptotics depending 
upon whether $\gamma$ is positive, negative
or zero.  Since the superpotential has a manifest symmetry under $\chi
\to -\chi$, we focus on $\chi >0$: If $\gamma$ is positive then
$\alpha \sim  \chi + {1 \over 2}\log({\gamma \over 2})$ for large (positive)
$\chi$.  If $\gamma$ is negative then $\chi$ limits to a finite value, $\chi_0$,
as $\alpha$ goes to $-\infty$.  If $\gamma =0$ then we get the
interesting ridge-line flow with $\alpha \equiv  -\chi$ all along the
flow.  Some of these flows are shown in Figure 1.    In all three cases, one
has $e^A \to 0$, and so the metric, \RGFmetric, becomes singular.

%%%%%%%%%%%%%%%%%%%%%%%%%%%%%%%
\goodbreak\midinsert
\vskip .5cm
\centerline{ {\epsfxsize 3in\epsfbox{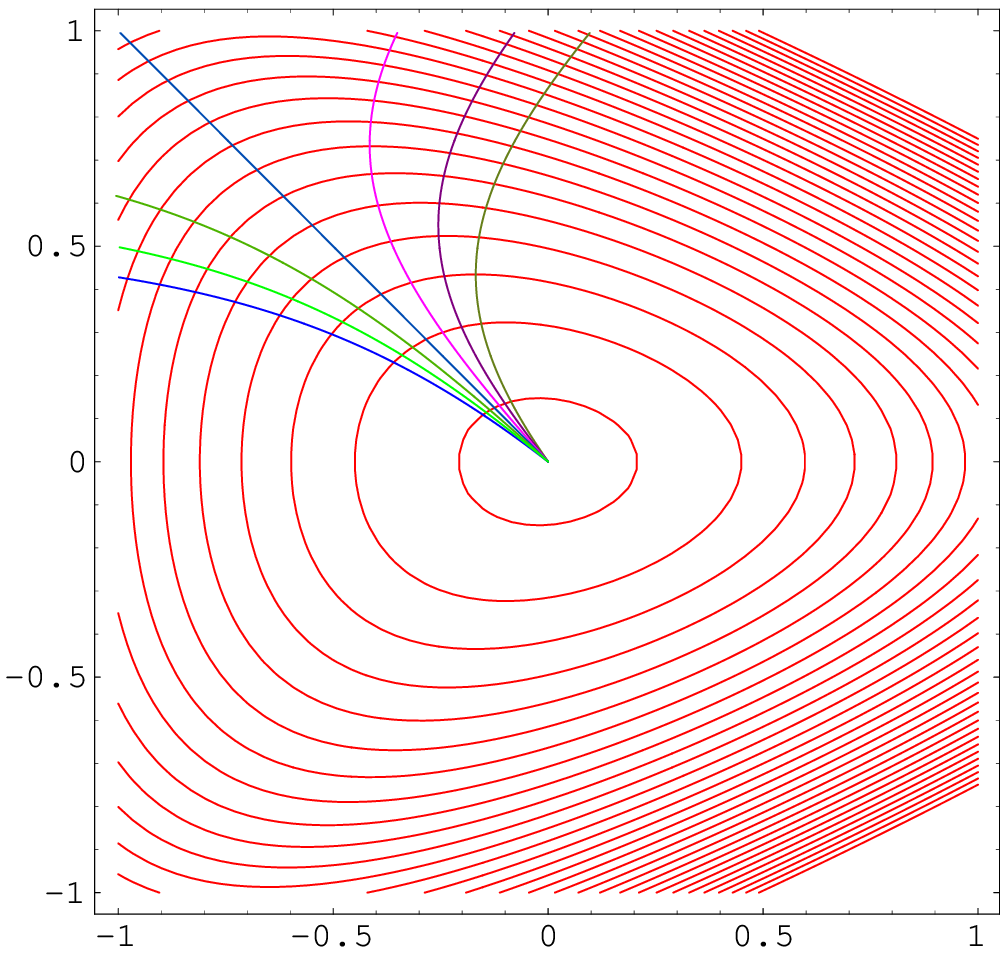}}}
\leftskip 2pc
\rightskip 2pc\noindent{\ninepoint\sl \baselineskip=8pt
{\bf Fig.~1}: Contours of the superpotential showing some of the
steepest descent flows. The horizontal and vertical axis are $\alpha$
and $\chi$, respectively. The ridge-line flow has $\gamma= 0$
and the three flows to the left and right have $\gamma< 0$ and
$\gamma>0$, respectively. To the right of the ridge the flows
asymptote to $\alpha \sim \chi$, on the ridge one has $\alpha = -\chi$,
and to the left of the ridge one has $\alpha \to -\infty$
as $\chi$ limits to a finite value.}
\endinsert
%%%%%%%%%%%%%%%%%%%%%%%%%%%%%%%

These results precisely parallel those of \PilchUE, and
suggest that the ``preferred flow,''  analogous 
to the purely massive Yang-Mills flow in four
dimensions, is the one with $\gamma=0$.   Following the parallel with the
$D3$-brane flow, the solution with $\gamma <0$ correspond to flows in
which the vevs of the operator $\cO_1$ dominates over the mass of the
fermions.  It thus approaches the Coulomb branch in the infra-red.  
The solutions with $\gamma >0$ are presumably unphysical as they
are for $D3$-branes. 
We will see later, the structure of the $M$-theory uplift
is very similar to that of $IIB$ uplift of the $D3$-brane flow \PilchUE.  
We will also see that our claim about the $\gamma=0$ flow is
supported by the $M2$-brane probes. 

\subsec{An Aside:  Continuous families of flow solutions} 

The supergravity potential \sugrpot\ exhibits a very curious feature that 
leads to a surprising conclusion in eleven dimensional geometry.  
In its original form, the gauged $\cN=8$ supergravity  theory has a local 
$SO(8) \times SU(8)$ symmetry.  The scalar manifold is $E_{7(7)}/SU(8)$,
and, as is usual, we have fixed the $SU(8)$ symmetry by 
chosing symmetric gauge in which all the scalars are represented 
by exponentials of the seventy  non-compact generators of $E_{7(7)}$.
There is still an $SU(8)$ {\it conjugation} action on this scalar manifold,
but it is  {\it only} the $SO(8)$ subgroup that is symmetry of the theory:  
Complex $SU(8)$ transformations map the four-dimensional
scalars into the pseudo-scalars and vice-versa.  In \Evars\ the supergravity
scalars and pseudo-scalars are represented by the real and imaginary
parts of the self-dual, complex forms, $\Sigma_{IJKL}$.   At the linearized
level, the scalars are internal metric perturbations while the
pseudo-scalars are internal components of the $3$-form, $A_{MNP}$.

It follows from our discussion above that the $U(1)$ rotations defined
by the phase angles $\phi$ and $\varphi$ in \polcoords\ are complex
$SU(8)$ rotations, and the real and imaginary parts of the $z_j$ are
scalars and pseudo-scalars respectively.   Turning on only the real part
of the $z_j$ represents a pure metric deformation, while turning
on the imaginary part represents turning on a tensor gauge field background,
which will then affect the metric at second order through the back-reaction.
Rotations of $\phi$ and $\varphi$ thus trade metric deformations 
for tensor gauge fields, and vice-versa.   The statement that such
$SU(8)$ rotations are not symmetries of the theory is a simply the
statement that such a rotation of the configurations will not 
generically preserve the Lagrangian, and thus will not be
consistent with the equations of motion.

However, the potential \sugrpot\ does {\it not} depend upon $\phi$ and
$\varphi$, and indeed both the potential and the kinetic term \scalkin\ are 
invariant under translations in $\phi$ and $\varphi$.  Thus, given a solution at with
some fixed value of $\phi$ and $\varphi$, we can trivially 
generate another by rotating the value of $\phi$ and $\varphi$.
In particular, the pure Coulomb branch solution with $\phi =\varphi=0$
can be rotated into a two parameter, continuous family of solutions
with non-trivial background fluxes.  Therefore, there is a family
of solutions that interpolates between the pure metric
 flow with $\phi =\varphi=0$, and the flow with \angchoice\ 
considered above.  

The flow with $\phi =\varphi=0$ has $\cW_1 = \cW_2 =\cW_3$,
and if we use the equations of motion \floweqs\ then this flow has sixteen 
supersymmetries.  This solution (with $\phi =\varphi=0$ has $\cW= \cW_1 = \cW_2 
=\cW_3$) actually solves the second order system of equations 
coming form \scalkin\ and \sugrpot\ for {\it any} fixed values 
of $\phi$ and $\varphi$.   However, the resulting flow will
not be supersymmetric for general values of $\phi$ and $\varphi$.
Partial supersymmetry will be
present in special cases:  For example, there will be eight supersymmetries
if $\phi=0$, and four supersymmetries if $\phi = \pm 2 \varphi$.

Thus the unexpected symmetry of the potential \sugrpot\  enables
us to generate continuous families of solutions of $M$-theory
in which metric deformations are traded for fluxes, and 
supersymmetry is partially, or totally broken within these 
families, with extra supersymmetries appearing on special flows.
A particularly simple, but illustrative example of this is presently being
studied in detail  \CNPNPW.

%%%%%%%%%%%%%%%%%%%%%%%%%%%%%%%%%%%%%%%%%%%%%%%%%%
\newsec{The flow solutions in $M$-theory}
%%%%%%%%%%%%%%%%%%%%%%%%%%%%%%%%%%%%%%%%%%%%%%%%%%

\subsec{The new solution}

From the large body of work on consistent truncation, and most particularly from
 \refs{\deWitNZ,\deWitMZ, \deWitIY},  we know that the solution given in the last 
section can be lifted to eleven dimensions. First, the metric for the eleven-dimensional 
solution can be obtained directly from the scalar field configuration
by employing the metric formula in   \deWitNZ. 
For the flows with \angchoice\ we find the following
frames:
\eqn\frames{\eqalign{  e^1 ~=~& \Omega\, e^A\, dt \,, \qquad
e^2 ~=~   \Omega\, e^A\, dx \,, \qquad e^3 ~=~  \Omega\, 
e^A\, dy \,, \qquad  e^4 ~=~  \Omega\,  dr \,, \cr
e^5 ~=~& 2\, L\, c^{-{1\over2}}\, \Omega\, d\theta  \,, 
\qquad e^6 ~=~   L \, \rho\, \Omega\, X_1^{-{1 \over2}}\,
\cos \theta \, \sigma_1 \,,  \qquad e^7 ~=~  L \, \rho\, 
\Omega\, X_1^{-{1 \over2}}\, \cos \theta \, \sigma_2  \,, \cr
\qquad e^8 ~=~ &  L \, \rho\,  \Omega\, (c\, X_2)^{-{1 \over2}}\, 
\cos \theta \, \sigma_3  \,,  \qquad e^9 ~=~ L \, \Omega\, 
X_2^{-{1 \over2}}\, \sin \theta \, \tau_1 \,, \cr
\qquad e^{10} ~=~ & L \,\Omega\, X_2^{-{1 \over2}}\, 
\sin \theta \, \tau_2 \,,\qquad e^{11} ~=~ L \, 
\Omega\, (c\, X_1)^{-{1 \over2}}\, \sin \theta \, \tau_3 \,,}}
where $c \equiv \cosh(2\,\chi)$ and $\rho=e^\alpha$.  
The functions $X_1, X_2$ and $\Omega$ are defined by: 
\eqn\fundefns{\eqalign{X_1 ~=~&  (\cos^2 \theta + \rho^2\, 
\cosh(2\,\chi) \, \sin^2 \theta)\,, \qquad X_2 ~=~ 
( \cosh(2\,\chi)\, \cos^2 \theta + 
\rho^2 \, \sin^2 \theta)\,,  \cr  \Omega~=~&
\big(\rho^{-2} \, \cosh(2\,\chi)\, X_1\, X_2  \big)^{1\over 6} \,,}}
The $\sigma_j$ and $\tau_j$ are independent sets of left-invariant 
one-forms on $SU(2)$, and satisfy $d\sigma_1 = \sigma_2 \wedge
\sigma_3$; $d\tau_1 = \tau_2 \wedge \tau_3$, plus cyclic permutations.
This metric has a manifest symmetry of $(SU(2) \times   U(1))^2$, 
and there is also an interchange symmetry:
\eqn\discsymm{\theta \to {\pi \over 2} - \theta \,, \qquad
\alpha \to - \alpha \,.}

The equations of motion (in the conventions of \refs{\PopeBD, \PopeJJ}) are:
\eqn\eqnmot{\eqalign{ R_{MN} ~+~ R \, g_{MN}  ~=~&  \coeff{1}{3}\, 
F^{(4)}_{MPQR}\, F^{(4)}_N{}^{PQR}\,,\cr 
\nabla_M F^{MNPQ} ~=~& -  \coeff{1}{ 576} \, \epsilon^{NPQR_1R_2
R_3R_4S_1S_2S_3S_4} \,F_{ R_1R_2
R_3R_4} \, F_{ S_1S_2 S_3S_4} \,.}}

Based upon previous experience \refs{\PilchUE,\PilchFU,\KhavaevYG}, 
we conjecture the following form for the $3$-form potential:
\eqn\Aansatz{\eqalign{A^{(3)} ~=~ & \widetilde W\, e^{3\,A} \, dt
\wedge dx \wedge dy ~+~ a_1(\chi,\theta)\, \sin \theta\, \cos \theta\,
d\theta \wedge \sigma_3 \wedge \tau_3 \cr & ~+~ a_2(\chi,\theta) 
\, \sin^2 \theta\, \cos^2 \theta\,
\sigma_1 \wedge \sigma_2 \wedge \tau_3 ~+~ a_3(\chi,\theta)\,
\sin^2 \theta\, \cos^2 \theta\,
\sigma_3 \wedge \tau_1\wedge \tau_2 \,.}}
The quantity, $\widetilde W$ is sometimes called
the geometric superpotential, and its form can be
inferred by techniques similar to those of \refs{\KhavaevYG,\CorradoNV}.
We indeed find that:
\eqn\geomW{\widetilde W ~=~ \coeff{1}{2} \, \rho^{-1}\, X_1 ~=~
\coeff{1}{2} \, \rho^{-1}\,(\cos^2 \theta + \rho^2\, \cosh(2\,\chi)
\, \sin^2 \theta)\, .}

We find that the Ricci tensor of the metric
obeys the identities:
\eqn\obvidens{R_{11} = - R_{22} = -R_{33}\,, \qquad
R_{66}=R_{77}\,, \qquad R_{99}= R_{10\,10} \,,}
and
\eqn\specidens{R_{66} + R_{88} = R_{11} \,, \qquad
R_{99} + R_{11\,11} = R_{11} \,.}
The identities \obvidens\ are a trivial consequence of
the symmetries of the metric, while the identities
\specidens\ are consistent with the special Anstatz,
\Aansatz, for the $3$-form.

To determine the functions $a_j(\chi,\theta)$, we proceed
much as in \PilchUE.  One computes the field strength, $F$, and
one finds that it has the form:
\eqn\Fform{\eqalign{F  ~=~ & h_1\,e^1 \wedge e^2 \wedge
e^3 \wedge e^4 ~+~ h_2\,e^1 \wedge e^2 \wedge
e^3 \wedge e^5 ~+~  b_1\,e^4 \wedge e^5 \wedge
e^8 \wedge e^{11} \cr & ~+~   b_2\,e^4 \wedge e^6 \wedge
e^7 \wedge e^{11} ~+~ b_3\,e^4 \wedge e^8 \wedge
e^9 \wedge e^{10} ~+~ b_4\,e^5 \wedge e^6 \wedge
e^7 \wedge e^{11}  \cr & ~+~  b_5\,e^5 \wedge e^8 \wedge
e^9 \wedge e^{10} ~+~ b_6\,e^6 \wedge e^7 \wedge
e^9 \wedge e^{10}  \,,}}
where the $h$'s and $b$'s  are given in terms of the functions
$\widetilde W$ and $a_j$.  One then computes the energy-momentum
tensor in terms of the $h$'s and $b$'s, and from this
one can immediately test, and verify that  \geomW\ yields
the correct geometric superpotential. One then looks at
differences of components of the energy momentum tensor,
and compares them with the corresponding differences 
between components the Ricci  tensor, using \geomW\ where
necessary.  The result is  typically an analytic expression
for the  difference of squares of two of the $b_k$.  One can
then factorize both sides and find the individual $b_k$.
The result is then integrated to get the $a_j$.
We find:
\eqn\ajresult{\eqalign{a_1(\chi,\theta) ~=~& L^3\, 
\tanh(2\,\chi)\,, \qquad a_2(\chi,\theta) ~=~ \coeff{1}{2}\,L^3\,
\rho^2 \, X_1^{-1} \sinh(2\,\chi)\,,\cr a_3(\chi,\theta) ~=~&
\coeff{1}{2}\,L^3\, X_2^{-1} \sinh(2\,\chi)\,.}}

A tedious, but straightforward calculation shows that
the $3$-form defined by \Aansatz\ and \ajresult\ satisfies
the generalized Maxwell equations in \eqnmot.
We thus have a complete ``lift'' of the supersymmetric flows
of section 2.

The results for both the metric and the potential are very similar
in form to the result for the corresponding $IIB$ flow given in
\PilchUE.  For the $IIB$ flow the metric on the $5$-sphere had the form:
\eqn\IIBmetr{\eqalign{
ds_5^2 & ~=~
{a^2\over 2} { (cX_1X_2)^{1/4}\over\rho^3} \left(
 c^{-1} d \theta^2 +\rho^6\cos^2\theta\,\Big({\sigma_1^2\over cX_2}
+{\sigma_2^2+\sigma_3^2\over X_1}\Big)+\sin^2\theta\, 
{d\phi^2\over X_2} \right) \,.
\cr } }
while the complex $2$-form potential was:
\eqn\IIBAanz{\eqalign{ A_{(2)} ~=~  e^{i\phi}\, \big( a_1(r,\theta)\,\cos\theta\,  
d\theta\wedge\sigma_1 ~+~  & a_2(r,\theta)\,\sin\theta\cos^2\theta\,\sigma_2 
\wedge\sigma_3 \cr ~+~  &  a_3(r,\theta) \,\sin\theta\cos^2\theta \, \sigma_1
\wedge d\phi   \big)\,,}}
with
\eqn\IIBsolutiona{\eqalign{
a_1(r,\theta)&~=~  -i\, L^2 \, \tanh(2\chi)\,, \qquad 
a_2(r,\theta) ~=~  i\,  L^2\,  \rho^6 \, X_1^{-1}\, \sinh(2\chi) \,,\cr
a_3(r,\theta)&~=~  L^2\, X_2^{-1} \, \sinh(2\chi) \, .\cr}
}
The functions, $X_j$, are very similar to those considered here, 
just with powers of  $\rho$ changed consistently throughout.  As one can see,
the forms are remarkably similar, and indeed, to collapse the eight-geometry
to the six-geometry one roughly drops the $\tau_3$ everywhere, and replaces
$\tau_1$ and $\tau_2$ by $d\phi$.  If one writes:
\eqn\oneforms{\eqalign{ \tau_1 ~\equiv~&  \cos(\phi_3)\, d\phi_1 ~+~ 
\sin(\phi_3)\, \sin(\phi_1)\, d \phi_2 \,, \cr
\tau_2 ~\equiv ~&  \sin(\phi_3)\, d\phi_1 ~-~ 
\cos(\phi_3)\, \sin(\phi_1)\, d \phi_2 \,, \cr
\tau_3 ~\equiv ~&  \cos(\phi_1)\, d\phi_2 ~+~   d \phi_3 \,.}}
then the foregoing suggests that one should identify $\phi$ with some 
combination of $\phi_1$ and $\phi_2$, while the other combination, 
which we will denote $\psi$ should be combined with $\phi_3$ to
get the compactifying circle for the  $M$-theory.  The remaining
combination of $\psi$ and $\phi_3$ will  then become the extra dimension 
that takes the $M2$-brane to the $D3$-brane.  One can, in fact track the
detailed identification through the gauged supergravities in four and
five dimensions.  We will not pursue this in detail here because
 the $T$-dual of the $IIB$ solution of \PilchUE\ is not exactly
the same as our $M$-theory solution:  The former  involves a smeared
distribution of $D3$-branes,  while the latter involves a $M2$-branes
that are localized in ${\Bbb R}^8$.

\subsec{Supersymmetries and orbifolds}

It is  very instructive to examine which of the geometric symmetries
act as $\cR$-symmetries.  Denote the  symmetries
of the metric and $3$-form by $SU(2)_\sigma \times U(1)_\sigma$
and $SU(2)_\tau \times U(1)_\tau$, where the subscripts denote
which of the left-invariant one-forms are involved.   While we have
not computed the supersymmetries in eleven dimensions explicitly,
one can use the results of section 3 and either the linearized
solution, or the metric uplift formula of \deWitNZ\ to relate the geometric
symmetries on $S^7$ to the symmetries acting in the four-dimensional
$\cN=8$ theory.  Doing this, we find that the unbroken supersymmetries are singlets
under $SU(2)_\tau \times U(1)_\sigma$, and that they transform as
${\bf 2}_{\pm 1}$ under $SU(2)_\sigma \times U(1)_\tau$.  One can check
that these transformation properties are consistent with these supersymmetry
parameters becoming the Weyl components of the $\cN=2$ supersymmetry
in the $IIB$ theory after a $T$-duality in one of the $\tau$-directions.

Finally, we note that because the supersymmetry parameters are inert
under $SU(2)_\tau$, the $\cN=4$ supersymmetry will be {\it preserved}
under any orbifold by a discrete subgroup of $SU(2)_\tau$.  Indeed, such
an orbifold construction can be used {\it ab initio} to reduce the supersymmetry
as in \refs{\DouglasSW,\JohnsonPY,\KachruYS}  to the half-maximal amount.
The flows considered here may thus be thought of the flows in the
untwisted sector of the resulting $\cN=4$ supergravity in four dimensions.
This closely parallels the corresponding discussion for gauged supergravity
in five dimensions (see, for example, \refs{\KhavaevFB, \CorradoWX}).

The bottom line is that the flows considered here may be thought of
as flows within the untwisted sectors of an $ADE$ family of orbifold
theories obtained by modding out the corresponding discrete group
of $SU(2)_\tau$.

%%%%%%%%%%%%%%%%%%%%%%%%%%%%%%%%%%%%%%%
\newsec{Brane distributions and probes}
%%%%%%%%%%%%%%%%%%%%%%%%%%%%%%%%%%%%%%%

\subsec{The harmonic distribution}

To understand the brane distribution in our solution, we first examine the 
simplest situation in which the fermion bilinears are set to zero.
A  flow that only involves the operator $\cO_1$ of \pertOps, is a pure Coulomb
branch flow, and it must preserve maximal supersymmetry.   In supergravity
such a flow lies  purely in the scalar sector, and must correspond to a standard
harmonic distribution of  branes.  Indeed, if one makes the further restriction 
of setting $\chi=0$ in  \polcoords\  then one must obtain an $SO(4) \times SO(4)$
 invariant distribution. This solution is easily mapped out, and correponds to a 
uniform distribution of branes on a solid $4$-ball.

To see this one simply sets $\chi =0$ in the equations of motion \floweqs.  The
eleven-dimensional metric becomes:  
\eqn\harmmet{\eqalign{ds_{11}^2 ~=~ H^{-2/3}  \, ( &  -dt^2 +   dx^2 + 
dy^2)    ~+~ H^{1/3}  \,   {4 \, L^2\over  \sinh \alpha}\,  \bigg( \, {X_0 \over 4\, \rho\, L^2}
\, dr^2   \cr ~+~ &   {X_0 \over \rho }\, d \theta^2  ~+ ~  \coeff{1}{4}\, \rho\,  \cos^2 \theta \,
\sum_{j=1}^3 \, \sigma_j^2 ~+~ \coeff{1}{4}\, \rho^{-1} \,  \sin^2 \theta\, 
\sum_{j=1}^3 \, \tau_j^2\, \bigg)\,.}}
where
\eqn\harmfn{ H~\equiv~  e^{- 3A} \,  \rho \, X_0^{-1} ~=~ \rho \,  \sinh^3 \alpha \,  X_0^{-1}  \,, 
\qquad X_0 ~\equiv~  X_1\big|_{\chi=0} ~=~ (\cos^2\theta ~+~ \rho^2\,
\sin^2\theta ) \,. }
 
By making a change of variables:
\eqn\newvars{  u~=~ {e^ \alpha  \over \sqrt{ (e^{2 \,\alpha} - 1)}} \, \cos \theta \,,
\quad v~=~ {1 \over \sqrt{ (e^{2 \,\alpha} - 1)}} \, \sin \theta \,,}
The metric reduces to the standard harmonic form:
\eqn\harmform{\eqalign{ds_{11}^2 ~=~ H^{-2/3}  \, ( &  -dt^2 +   dx^2 + 
dy^2)  \cr    + ~& 8\, H^{1/3} \, L^2\, \bigg( du^2 +  dv^2 + \coeff{1}{4}\, u^2 \,
\sum_{j=1}^3 \, \sigma_j^2 + \coeff{1}{4}\, v^2 \, \sum_{j=1}^3 \, \tau_j^2 \bigg)\,,}}
where the metric in the second set of parentheses is the  flat
metric on $\IR^8$.  One can then check that in these new coordinates, the
harmonic function, $H(u,v)$, of \harmfn\  is given by:
\eqn\Gfnform{H(u,v)~=~  {\rm const. }  \, \int_{z^2 <1} \, { d^4z \over 
(u^2  ~+~ (\vec v - \vec z)^2)^3 } \,,}
where $\vec v$ and $\vec z$ are vectors in $\IR^4$.
This means that the $M2$ branes are spread out into a solid $4$-ball,
defined by $v^2 <1$,  with a constant density of branes throughout the ball.

\subsec{The general distribution: Brane probes}

One can attempt to find the general distribution of branes by naively
trying to force the general eleven-metric into a the ``harmonic form''
\harmmet.  That is, one now writes the general metric in the form:
\eqn\forcedform{ ds_{11}^2 ~=~ H^{-2/3}  \, ds_{2,1}^2    ~+~ H^{1/3} \, ds_{8}^2   \,,}
where 
\eqn\Hfunction{H ~\equiv~  e^{- 3A} \,  \Omega^{-3} \,,}
and
\eqn\transmet{\eqalign{ds_8^2 ~\equiv~ &    \Omega^3 \, e^A \, L^2 \,
\Big[ \, {1 \over L^2}\,  dr^2 ~+~  {4 \over c}\,  d{\theta}^2 ~ +~ {{\rho}^2 \over X_1} \,  \cos^2
\theta\, (\sigma_1^2  +  \sigma_2^2) ~+~  {{\rho}^2 \over c \,  X_2} \, \cos^2\theta \, 
\sigma_3^2   \cr& \qquad \qquad \quad  ~+~  {1  \over   \,  X_2}   \sin^2\theta \,
(\tau_1^2  +  \tau_2^2) ~+~ {1 \over c \,  X_1} \, \sin^2\theta \, \tau_3^2\, \Big] \,,}}
These expressions are only a formal parallel, and the function $H$
is certainly not harmonic, but $H$ should encode the brane distribution.
The way to do this correctly is, of course, to use brane probes.

The action for the membrane probe has the following pieces:
\eqn\BPact{S 
~= ~I^{DBI}~ + ~I^{WZ}  ~\equiv~ \mu_2~  \int \, d^3 \sigma\, \Big[  -
\sqrt{ -  det(\tilde g)}  ~+~ \coeff{1}{3}\, \tilde A^{(3)}\Big] \,.}
where $\tilde g$ and $\tilde A^{(3)}$ denote the pull-back of the
metric and the $3$-form onto the membrane.  The  normalization of the
$A^{(3)}$-term in \BPact\ is twice the usual normalization since
this is the normalization that we have used in the eleven-dimensional
equations of motion. 

Following the usual  approach, we take the probe to be
parallel to the source membranes, and assume that
it is traveling at a small velocity transverse to its world-volume.
Expanding to second order in velocities, one obtains a kinetic
energy and a potential term:
\eqn\probeexp{\eqalign{I^{probe}  ~\equiv~ & \int~dt~\big( T ~-~ V\big)\cr ~=~ 
&\coeff{1}{2}~m_{M2} \,   \int dt\   \Big( G_{m\,n}\, v^m\, v^n ~-~ e^{3 A}\, 
\big({\Omega}^3~-~  2\,  \widetilde W \big) \Big) \,\,,}}
where we have replaced the $\mu_2$ and the integral over the spatial
volume of the probe by  the mass $m_{M2}$.  In this expression, the indices
$m,n =1,\dots,8$ run over the directions transverse to the brane, and the metric
$G_{mn}$ is precisely that of \transmet:
\eqn\probemet{ ds_8^2 ~\equiv~   G_{mn} \, dx^m\, dx^n ~=~ 
\Omega\, e^A \, \Big(\, g_{mn} \, dx^m\, dx^n\,\Big) \,,}
where the metric $g_{mn}$ is obtained from the last eight frames in  \frames.

The potential seen by the probe membrane is thus:
\eqn\probepot{ V~=~m_{M2}\, e^{3A}\, \big(\, {\Omega}^3~-~2\, \widetilde W\,  
\big) ~=~\coeff{1}{2}\, m_{M2}\,   { k^3 \, \rho^2 \over  \sinh^3 2\, \chi } \,  \sqrt {X_1}\,  
 \big(\, \sqrt{c\, X_2}~-~\sqrt{X_1} \, \big) \,.}
This is very similar to the result found in \refs{\BuchelCN,\EvansCT}.

The force-free regions, or moduli spaces of the brane probe, are given
by the loci where the potential vanishes.    From \probepot\  one sees that
there are two such  regions:
\item{{\bf I}}: \quad  $ \sqrt{c\, X_2}~=~\sqrt{X_1} ~~\Leftrightarrow~~ \cos\theta~=~0$ .
\item{{\bf II}}: \quad    $\rho ~=~0$ .

Recall that the harmonic branes were spread in the $v$-direction, which
corresponds to $ \cos\theta~=~0$, and so this is going to be the
physically interesting direction.  We will therefore briefly consider locus II, 
and study locus I in more detail.

\subsec{The special loci}

Locus II  only exists for the flows with $\gamma <0$  in \alsoln.  Let  $\chi_0$ be the
value of $\chi$ at which $\rho$ vanishes, and define 
$c_0 = \cosh(2\chi_0)$, $s_0 = \sinh(2\chi_0)$.   As one $\rho \to 0$, one has,
from \floweqs\ and \superpot:
\eqn\asymr{ dr ~\sim 2\,L\, d\rho \,.}
The metric, $ds_8^2$, becomes, to leading order in $\rho$:
\eqn\metlim{\eqalign{ ds_8^2 ~\sim~ \coeff{4\, k \, L^2}{s_0} \,  \big[ c_0 
\, \big( \,  du^2 ~+~&\coeff{1}{4}\,  u^2  (\sigma_1^2 + \sigma_2^2 + \coeff{1}{c_0^2}
\, \sigma_3^2\,) \big)\cr & ~+~ dw^2 ~+~ \coeff{1}{4}\, w^2\, (\tau_1^2 + \tau_2^2 + \tau_3^2 ) 
\, \big] \,,}}
where $u \equiv \rho \cos\theta$ and $w\equiv \sin \theta$.  Note that to leading
order, $w$ does not involve $\rho$, and   thus $w$  remains finite in the
range $|w| \le 1$ as $\rho \to 0$.
In this same limit, the brane-distribution  function, \Hfunction, is, to leading order:
\eqn\Hfnres{H ~\sim~   {s^3 \over c}\, {1 \over u^2} \,.}
This is once again consistent with a uniform distribution of $M_2$-branes spread in a  
four-dimensional  ball.   At $\rho =0$ the branes thus see the completely flat moduli space,
parametrized by $w$.  

This result is not too surprising since the $\gamma<0$ solutions correspond to
a solutions in which the Coulomb branch vevs of the scalars are dominating
over the fermion mass terms, and so in the infra-red, the probe brane should
see the moduli space of the original Coulomb branch. 
 
The physically most interesting flows are those with $\gamma =0$.  These
are the ones that should describe the purely massive flows with no Coulomb
branch component to the flow.    For such flows locus II is singular, and there 
only remains locus I.  The moduli space is particularly interesting.  

Setting $\theta = {\pi \over 2}$ in  \transmet\   yields the $4$-metric:
\eqn\degentwo{ds_4^2 ~=~  {c \over s^3} \, d\chi^2 ~+~ 
{c \over 4 \, s } \,(\tau_1^2 + \tau_2^2) ~+~  {1 \over 4\, c\, s} \,   \tau_3^2  \,,}
where $c = \cosh(2\chi)$, $s = \sinh(2\chi)$.  There are several remarkable
features to note immediately.  First, $\rho$ has cancelled out completely,
and so this limiting metric is independent of the choice of $\gamma$.
Secondly, as $\chi \to \infty$ the $2$-sphere defined by $\tau_1$
and $\tau_2$ limits a sphere of radius ${1 \over 2}$.  Indeed, define a new radial
variable, $\mu = \sqrt{c \over s}$ and one finds:
\eqn\EHmet{ds_4^2 ~=~  { d\mu^2  \over \big(1 - {1 \over \mu^4} \big)}   ~+~ 
\coeff{1}{4} \,\mu^2 \,(\tau_1^2 + \tau_2^2) ~+~\coeff{1}{4} \, \mu^2  \,  
\Big(1 - {1 \over \mu^4} \Big) \, \tau_3^2  \,,}
which is  almost exactly the Eguchi-Hanson metric.  It is thus the 
Ricci-flat, 
hyper-K\"ahler metric on the blown-up $A_1$ singularity.  The blown-up $2$-sphere, 
defined by $\tau_1$ and $\tau_2$, has radius ${1 \over 2}$.

It is not exactly the Eguchi-Hanson metric because the periodicity of the angular
coordinate in $\tau_3$ is not correct.  If one uses \oneforms\ then the angle
$\phi_3$ has period $4\,\pi$, but for \EHmet\ to be regular as $\mu \to 1$, 
$\phi_3$ must have a period of $2\,\pi$.  Thus, at large $\mu$, the surfaces of
constant $\mu$  in the  Eguchi-Hanson metric must look like $S^3/{\Bbb Z}_2$ 
and not like $S^3$.  Our flow solution
started with $S^7$ without any discrete identifications, and so the 
$4$-slice represented by \degentwo\ does not have the ${\Bbb Z}_2$
identifications needed to get Eguchi-Hanson exactly.

On the other hand, we noted at the end of section 4 that our flow
solution and its supersymmetries are invariant under $SU(2)_\tau$,
and so may be modded out by any discrete subgroup $SU(2)_\tau$.  Therefore we  can
trivially modify our solution to accommodate the asymptotics of any ALE space, and 
indeed if we go to the ${\Bbb Z}_2$ orbifold then we get the $M2$-brane moduli
space to be Eguchi-Hanson on the nose.  It would be interesting to try to generalize
our solution to get an arbitrary ALE space as the moduli space of the
$M2$-branes.

Thus the $M2$-branes see, at least locally, a classic hyper-K\"ahler moduli space, 
and as $\chi \to \infty$, or $\mu \to 1$, the branes are spread out over a finite-sized
$2$-sphere.   Indeed, for the $\gamma =0$ flow, the brane-distribution function, \Hfunction,
exhibits the proper asymptotics.  First observe that the coordinates $\chi$ and
$\mu$ are singular as $\chi \to \infty$.  Indeed, $ds_4^2  \sim  4 e^{- 4\,\chi} \, d\chi^2
+ \dots$, which means that a good radial coordinate is $R \equiv e^{- 2\,\chi}$.
Then, for   $\theta = {\pi \over 2}$, and $\gamma =0$, the brane-distribution function 
\Hfunction\ becomes:
\eqn\Hfnres{H ~=~ {\sinh^3(2\,\chi) \over \rho^{4} \,  \cosh(2\,\chi)} ~\sim~
 {1\over 4}\,  e^{8\,\chi} ~\sim~  {1\over 4} \,   {1 \over R^4} \,,}
which is precisely the correct asymptotics for a uniform distribution of
$M2$-branes spread over a $2$-sphere at $R=0$.   Moreover, for 
$\theta ={\pi \over 2}$ the $3$-form potential becomes:
\eqn\simpA{A^{(3)} ~\sim~  H^{-1} \, dt \wedge dx \wedge dy \,,}
which further supports our interpretation.

It is worth recalling the parallel result for the $D3$-brane flow of  
\refs{\BuchelCN,\EvansCT}.
For $D3$-branes the corresponding Coulomb branch involved
spreading the $D3$-branes uniformly over a two-dimensional disk.
Turning on a fermion mass involved turning on background 
$2$-form gauge fields, and a non-trivial dilaton and axion.  The $IIB$  background
could thus be thought of as a dielectric mix of additional $5$-branes and
$7$-branes.   Using $D3$-brane probes on locus II involved approaching
the solution in the two-dimensional direction ($\theta ={\pi \over 2}$) where
the disk of branes had spread on the pure Coulomb flow.   In the more
general background, with non-zero fermion masses, the $2$-form fields
vanished on locus II, but the dilaton and axion remained non-trivial, and
could be used to track the distribution of $D3$-branes. (In a sense, they
had dissolved in the dielectric $7$-branes that sourced the dilaton and axion.)
In particular, the  $D3$-brane probe of the  ``purely massive''  ($\gamma =0$)
flow was studied extensively in  \refs{\BuchelCN,\EvansCT}.  
It was shown that the ``Coulomb disk'' had 
flattened to a line distribution of branes in which the brane density was no longer 
constant, but seemed to ``remember'' its disk-like origins:
\eqn\branedensity{\rho(y) ~=~ \sqrt{a_0^2 - y^2}\,.}

For the $M2$-brane flow we have found that the four-ball of the
Coulomb branch has collapsed, and then been blown up into a 
$2$-sphere with an apparently uniform $M2$-brane distribution.

\newsec{Conclusions}

Our motivation in doing this work was to try to get a broader geometric
understanding of supersymmetric solutions in the presence of $RR$-fluxes.
We ultimately hope to find a geometric characterization of general 
supersymmetric holographic flows, and go beyond those that we
can access using lower-dimensional gauged supergravity.

In this paper we focused on flow solutions with eight supersymmetries
in both $F$-theory and $M$-theory.  These solutions involve a 
transverse eight-manifold with half-dimensional forms ({\it i.e.} $4$-forms)
only.  We believe that such configurations, along with such a high level of 
supersymmetry should be more amenable to classification than, say, their
$IIB$ cousins with a transverse six-manifold and $1-$, $3-$ and $5$-form
fluxes.    Within $M$-theory, if there is a moduli space for the  brane probe 
then a hyper-K\"ahler structure is  guaranteed by the eight supersymmetries.
Within $IIB$ supergravity, eight supersymmetries means $\cN=2$ supersymmetry,
which only implies a K\"ahler moduli space.  However, we showed that  lifting the 
 $IIB$ flows to $F$-theory extends this  to a four-dimensional hyper-K\"ahler 
moduli space whose singularities coincide with the enhan\c con. 
 
Our explicit $M$-theory flow was constructed to precisely parallel the 
$\cN=2$ supersymmetric flow in the $IIB$ theory.  Given that we started
with a topologically trivial transverse space, with no orbifold identifications,
we had anticipated finding a Coulomb moduli-space that was topologically 
${\Bbb R}^4$.  Instead, we found that it was Eguchi-Hanson {\it without}
the proper modding out by ${\Bbb Z}_2$, but with the branes  smoothly
distributed over the non-trivial $S^2$.  In addition, we noted that we could
mod-out our solution so as to preserve the supersymmetry and yet make
this Coulomb moduli space have the large radius asymptotics appropriate to
any ALE space:  $S^3/\Gamma$ where $\Gamma$ is any discrete subgroup 
of $SU(2)$.  In particular we can mod out the entire solution by ${\Bbb Z}_2$
so that it still has eight supersymmetries and has exactly the Eguchi-Hanson
moduli space.

While  the example presented here is far simpler than that of \PilchUE,
we are still not able to fully characterize the flow geometries with eight
supersymmetries.  However, the results presented here provide what appears 
to be a very fruitful line of attack that is very much in the same spirit as using 
$G$-structures (see, for example, \GauntlettFZ).   First, there is the presence of the 
four-dimensional 
hyper-K\"ahler moduli space on the residual Coulomb branch.   While this
is very constraining, we would obviously like to understand how  this 
geometry is extended and generalized in directions transverse to the moduli space.
It turns out that the eleven-dimensional supersymmetries will tell us precisely
how to achieve this.   To be more precise,  this solution has eight
supersymmetries contained in four supersymmetry parameters, $ \epsilon_{(i)}$.
The group transformation property of these spinors under $SU(2)_\tau \times 
U(1)_\tau$ is precisely consistent with these spinors being made from tensor
products that involve the covariantly constant spinors on the Eguchi-Hanson
space.  Indeed, this is also required for consistency with the supersymmetry 
transformations on the probe brane.   Now consider the  bilinears:
\eqn\bilins{J^{(ij)}_{\mu \nu} ~\equiv~  \bar \epsilon_{(i)} \gamma_\mu   
\gamma_\nu \epsilon_{(j)}\,.}
Because the $ \epsilon_{(i)}$ satisfy $\delta \psi_\mu =0$,  the $J^{(ij)}$  must
be $2$-forms that satisfy first-order differential equations.   In the direction 
of the Eguchi-Hanson space  these spinors must be covariantly constant,
and these  bilinears therefore yield the hyper-K\"ahler forms.  Thus
by finding the Killing spinors of the complete solution we will be able to explicitly 
see how the hyper-K\"ahler structure is extended away from the Eguchi-Hanson slice.
We are currently pursuing this idea, and computing the explicit supersymmetries.

There is one other observation that suggests a special geometry transverse
to the brane.  In section 3 we made a choice  \superpot\ of superpotential.
We could equally well have made the opposite choice: 
$ \cW = \cW_2 = \cW_3$, which amounts to sending $\alpha \to - \alpha$
in our solution.  In section 4 we noted in \discsymm\ that this was a symmetry
of the full solution when coupled with the rotation $\theta \to {\pi \over 2} - \theta$.
This means that a flow that uses the other choice of superpotential will
generate another Eguchi-Hanson slice at $\theta = 0$, which is exactly transverse
to the Eguchi-Hanson slice described above.   So a relatively trivial change in 
the equations  of motion creates another hyper-K\"ahler slice transverse to the original
one.  There will then be a directly parallel story with the spinors, and the extension
of the hyper-K\"ahler forms into the whole eight-manifold.   By combining all
the $2$-forms generated in this manner, we hope to be able to pin down the geometry
very precisely.

\bigskip

\bigskip
\leftline{\bf Acknowledgements}

NW  would like to thank the {\it Isaac Newton Institute} in Cambridge, UK for
its hospitality while some of this work was done.  He would also like to
thank J.~Gauntlett for helpful conversations.  
This work was supported in part by funds
provided by the DOE under grant number DE-FG03-84ER-40168.

%%%%%%%%%%%%%%%%%%%%%%%%%%%%%%%%%%%%%%%%%
% End
%%%%%%%%%%%%%%%%%%%%%%%%%%%%%%%%%%%%%%%%%
\listrefs
\vfill
\eject
\end